# A model for vortex formation in magnetic nanodots

S. A. Leonel,[a)] I. A. Marques, and P. Z. Coura
*Departamento de Física, ICE, UFJF, Juiz de Fora, MG 36036-0900, Brazil*

B. V. Costa
*Laboratório de Simulação, Departamento de Física, ICEX, UFMG, 30123-970 Belo Horizonte, MG 36036-0900, Brazil*

We use Monte Carlo simulation to study the vortex nucleation on magnetic nanodots at low temperature. In our simulations, we have considered a simple microscopic two-dimensional anisotropic Heisenberg model with term to describe the anisotropy due to the presence of the nanodot edge. We have considered the thickness of the edge, which was not considered in previous works, introducing a term that controls the energy associated to the edge. Our results clearly show that the thickness of the edge has a considerable influence in the vortex nucleation on magnetic nanodots. We have obtained the hysteresis curve for several values of the surface anisotropy and skin depth parameter ($\xi$). The results are in excellent agreement with experimental data. © *2007 American Institute of Physics*. [DOI: 10.1063/1.2809408]

## I. INTRODUCTION

Recent advances in the technology for fabrication of nanoscale devices[1] have stimulated a great interest in the properties of submicron sized patterned magnetic elements.[2,3] The properties of nanostructured magnets, with size comparable to or smaller than the ferromagnetic domain size, offer a great potential for new physics. Their expected applications include magnetic random access memory, high-density magnetic recording media, and magnetic sensors. A special case is the possibility of using the magnetic quantum dot as a bit element in nanoscale memory devices. Magnetic circular nanostructures that exhibit a curling magnetic configuration, named vortex, are considered the basic structure for magnetoelectronic random access memory. By a vortex we mean a special configuration of spins similar to the stream lines of a circulating flow in a fluid. A precise definition of a vortex[4] will be given below. Direct experimental evidence for the existence of magnetic vortex states has been found by magnetic force microscopy (MFM). Shinjo *et al.*[5] have used MFM to characterize magnetic nanodots of permalloy (NiFe) with a thickness of 50 nm and radius between 300 nm and 1 $\mu$m. Spin-polarized scanning tunneling microscopy and the direct observation of the magnetization distribution in nanoscale iron islands with magnetic vortex cores have been reported.[6] Lorentz transmission electron microscopy allows *in situ* magnetizing experiments with thin samples, and it has been used to characterize the magnetization distribution in a number of different geometrical arrangements.[7] Cowburn *et al.*[8] have measured the magnetic properties of nanomagnets fabricated by high-resolution electron beam lithography. By applying a magnetic field in the plane of the nanomagnet they found that as the applied field is reduced from minus saturation, the nanomagnet retains full magnetization, until a critical field slightly below zero, at which point nearly all magnetization is lost. The magnetization then progressively reappears as the field is increased from zero, until positive saturation is achieved. The sudden loss of magnetization close to zero field is believed to be the signature of a vortex phase in the system.[8–11]

As long as one could manipulate the vortex states other possibilities would emerge. In fact, one way toward this control is obtained by removing some small portions of the magnetic nanodot, in such a way that the cavities so created work by attracting and eventually pinning the vortex around themselves.[12–16]

Further theoretical studies focusing on the ground state magnetic configuration in nanodisks have mostly been Monte Carlo simulations.[17–20] They have confirmed that a vortexlike structure could be responsible for the observed experimental behavior of magnetic nanodots. Kireev and Yvanov[21] using a discrete microscopic Hamiltonian approach introduced by Kodama *et al.*,[22] have shown that such a model can support vortexlike excitations. However, the authors argued that the configurations with structures they named half vortices, pinned at opposite sides of the border of the nanodot, are the more stable configurations at strong surface and easy-plane anisotropy. Although the theoretical models were able to describe some important features of the experimentally observed properties of vortex developed in nanomagnetic structures, in all cases the vortex has to be introduced by hand in the system.

In this work we show how possible is it to judiciously modify the model discussed in Refs. 21 and 22, so that it has a unique vortex in its ground state. Besides, the vortex is shown to be stable against application of an external magnetic field. Our results show that the model recover the main thermodynamical properties experimentally observed in magnetic nanodots structures. This paper is organized as follows. In Sec. II we present an operational definition of the vortex and introduce the model we are going to work with. In Sec. III we show some results obtained from a numerical

---

[a)]Electronic mail: sidiney@fisica.ufjf.br

approach we have used to simulate the model. In Sec. IV we compare our results with some experimental measurements.

## II. THEORETICAL BACKGROUND: MODEL FOR NANOPARTICLES

We can consider a magnetic nanodots represented by a small cylinder of radius $r$ and thickness $L$, so that its aspect ratio $L/r \ll 1$. For fine magnetic nanoparticles the role of the surface becomes much more important than for bulk materials and a two-dimensional (2D) model can be used to represent a flat nanodot. Soon, a microscopic model to describe a spin distribution inside magnetic nanoparticles is based on a discrete spin Heisenberg Hamiltonian defined over a lattice by the following 2D anisotropic model.[19–24]

$$H = -J \sum_{\langle ij \rangle} (S_i^x S_j^x + S_i^y S_j^y + \lambda S_i^z S_j^z) + B \sum_i^* (\mathbf{S}_i \cdot \mathbf{n}_i)^2. \quad (1)$$

Here $\mathbf{S}_i$ is a classical three-component spin variable defined on the site ($|\mathbf{S}_i|=1$) and $J>0$ stands for the exchange integral. $\lambda$ is an exchange anisotropy parameter. $B$ describes the anisotropy due to the presence of the edge where $\mathbf{n}_i$ is a vector normal to the edge. The first sum is over first neighbors and the second is over the last layer in the lateral edge. The last term in Eq. (1) represents a strong enough single-ion surface border anisotropy. Although it cannot be considered as the actual magnetostatic energy, we expect that it could imitate its role, say, increasing the total energy as long as the spins develop normal components at the borders of the nanodots (see Ref. 21 and references therein). Indeed, this anisotropy has the effect of keeping the original cylindrical-like profile of the vortex near these borders, although allowing it to deform in other regions along the nanodot face. The properties of the model in the thermodynamical limit are well established. It is important for us to know that the model can support nonlinear excitations named vortices. The vortex structure can be obtained in the following way. In the thermodynamical limit the edge term plays no rule and can be dropped out. Following Ref. 25 we can write a continuum version of Eq. (1) as

$$H \approx \frac{J}{2} \int \left[ \left(1 - \frac{\delta}{2} \cos^2 \theta \right)(\nabla \theta)^2 + \sin^2 \theta (\nabla \phi)^2 + \delta \cos^2 \theta \right], \quad (2)$$

where $\delta = 1 - (\lambda/2)$. The spin components were parametrized by using spherical angles

$$\mathbf{S} = \{\sin \theta \cos \phi, \sin \theta \sin \phi, \cos \theta\}. \quad (3)$$

By minimizing Eq. (2) in relation to angles $\theta$ and $\phi$ we obtain

$$\nabla^2 \phi = 0, \quad (0 < \phi < 2\pi), \quad (4)$$

$$\left(1 - \frac{\delta}{2}\cos^2\theta\right)\left(\frac{d^2\theta}{dr^2} + \frac{1}{r}\frac{d\theta}{dr}\right) + \frac{\delta \sin 2\theta}{4}\left(\frac{d\theta}{dr}\right)^2 - \frac{\sin 2\theta}{2r^2}$$

$$+ \frac{\delta}{2}\sin 2\theta = 0. \quad (5)$$

It is easy to check that $\phi = \pm \arctan(y/x)$ is a solution for the first equation together with ferromagnetic boundary conditions. This kind of solution is named as a vortex $(+)$ or antivortex $(-)$. The equation for the out-of-plane spin component $\theta$ has asymptotic solutions,

$$\theta \simeq \begin{cases} \dfrac{\pi}{2} - ae^{-\sqrt{\delta}r}, & r \to \infty \\ br^{[1-(\delta/2)]^{-1/2}}, & r \to 0. \end{cases}$$

A characteristic length scale is provided by $1/\sqrt{\delta}$ which can be interpreted as the vortex core. The energy of a single vortex can be estimated by using the solutions above,

$$H \approx \frac{J}{2} \int_a^R \sin^2 \theta (\nabla \phi)^2 \approx \pi J \left[ \ln\left(\frac{R}{a}\right) - \int_a^R \frac{\pi}{2}\frac{e^{-\sqrt{\delta}r}}{r} dr \right]. \quad (6)$$

Here $R$ is the vortex size and $a$ is an infrared cutoff, normally taken as the lattice size. The vortex energy diverges with diverging size. If $\lambda = 1$ ($\delta = 0$), the model turns out to be the isotropic Heisenberg model and the vortex becomes an instanton with finite energy. A more complete study on the 2D anisotropic Heisenberg model can be found in Refs. 4 and 26 and references therein.

A magnetic nanodot is a finite magnetic system, and the edge produces an anisotropy in the material. In that case we have to treat the Hamiltonian with the edge term. Kireev and Ivanov,[21] considering a model with a purely planar spin distribution ($\lambda = 0$, the $XY$ Heisenberg model) to represent a magnetic nanodot, have shown that in the continuous limit a vortex solution emerges from Eq. (1) as well as a capacitor-like solution they named two *half vortex* excitation (it must not be confused with the true half vortex as discussed by Lee and Grinstein[27]) with energies

$$E_{\text{vortex}} = \pi J \ln \frac{R}{a},$$

$$E_{\text{half}} = \pi J \left( \ln \frac{R}{a} - \ln 2 \right). \quad (7)$$

It is clear that in this limit the capacitor configuration is energetically preferable than the vortex solution. However, experimental evidences point out in the direction of the existence of very stable low energy vortex states at finite temperature.[5–11]

A consistent model to describe the vortex in nanomagnets must reproduce the experimental data. The Hamiltonian model defined in Eq. (1) seems to have the main ingredients to have a vortex as a stable solution of the equations of motion of the system. However, that solution does not correspond to the lower energy excitation being unstable at finite temperature. The boundary condition contained in the definition of Hamiltonian (1) is essential to the appearing of the vortex solution. In the thermodynamic limit that bias is not needed, since vortex always appears accompanied by an antivortex so that the energy of the pair is finite.

To judiciously define a model that can describe a stable vortex in a magnetic nanodot, we start by noting that the definition of the edge thickness is a point that is not clear. We know, for example, that the edge effects can be detected up to several layers inside the material (see, for example, Ref. 28 and references therein). In the thermodynamic limit (infinite volume) the edge term is not important, however, for finite systems it defines the geometry of the ground state. The introduction of a special field that acts only in the spins at the edge requires a more precise definition of what we mean by edge, which cannot be easily done. A model that preserves the boundary condition and is much more workable retaining the symmetry and the edge effects inside the material can be constructed by introducing a decaying field at the boundary. The edge energy being controlled by a *edge depth* ($\xi$) as defined below.

$$H = -J \sum_{\langle ij \rangle} (S_i^x S_j^x + S_i^y S_j^y + \lambda S_i^z S_j^z) + B \sum_i e^{-D_i/\xi} (\mathbf{S}_i \cdot \mathbf{n}_i)^2. \quad (8)$$

Now all sums run over all sites in the system. The term $e^{-D_i/\xi}$ controls the thickness of the edge. Here $D_i$ is the distance between the last layer in the lateral edge and the site $i$, to be taken along the $\mathbf{n}_i$ direction. The model defined by Eq. (1) is recovered in the limit of small $\xi$. In the next section we present some numerical results that clearly show that the most stable ground state solution for $\lambda=0$ is a vortex state for finite $\xi$. We have simulated the model [Eq. (8) with $\lambda=0$] in several situations. By applying an in-plane magnetic field we were able to recover, in detail, the hysteresis curves obtained in several experiments. The simulation does not need the introduction of any *ad hoc* condition.

## III. COMPUTATIONAL METHODS AND RESULTS

We want to simulate the Hamiltonian [Eq. (8) with $\lambda=0$] we have introduced in the last section. Our simulations are done through a Monte Carlo (MC) method using the METROPOLIS algorithm.[29] In our simulations we used the temperature in units of exchange integral $J$. For each temperature and surface anisotropy $B$, we obtain the most stable spin configuration. Each point in our simulation is the result of the average of $10^6$ independent configuration. The error bars are smaller than the symbols when not indicated. Initial spin configurations were taken as random. Figures 1(a) and 1(b) show typical ground state equilibrium spin configurations in nanodots of radius $r=20a$, obtained in our MC simulations. These configurations are in agreement with the theoretical results of Kireev and Ivanov.[21] Distance is measured in units of the lattice parameter $a$. In Figs. 2(a)–2(f) we show a lot of plots of the energy of the vortex and capacitor configurations for several values of the surface anisotropy and skin depth ($\xi$), as a function of the nanodot size $r/a$, at temperature $T/J=0.1$. The simulation results show that for the ground state configuration (1) the larger the skin depth ($\xi$), the more stable the vortex structure is, and (2) a large surface anisotropy ($B$) lowers the vortex configuration energy.

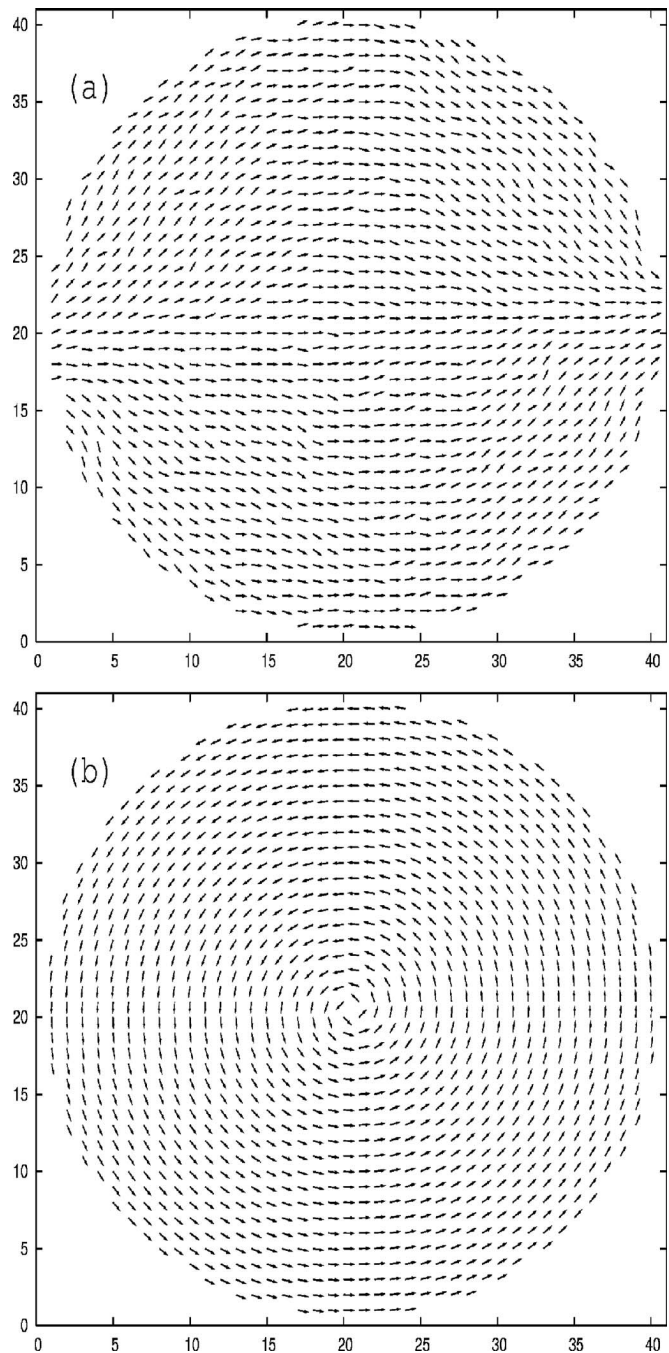

FIG. 1. Typical equilibrium configurations of a magnetic nanodot: (a) a capacitor (two half vortices) configuration; (b) a vortex configuration.

In order to compare our model with the experimental data, we introduce an external magnetic field, $h_{\text{ext}}=H^y/J$ in the $y$ direction, so that it contributes to the energy with the term $-H^y \sum_i S_i^y$. We vary the external field in the interval $-0.6 \leq h_{\text{ext}} \leq 0.6$ in steps of size $\delta h_{\text{ext}}/J=0.03$. To equilibrate the system at each field we use $10^6$ MC step. We start the simulation at low field $h_{\text{ext}}=-0.6$ and fixed temperature $T/J$. The field is increased up to $h_{\text{ext}}=0.6$, and then is lowered further down to $h_{ext}=-0.6$. In Fig. 3(a) we show our MC simulation results for the hysteresis curve for a nanodot with $r/a=20$, $B/J=3.0$, and $\xi=1$. In this case we do not observe the presence of the central plateau. This is a consequence of the absence of a vortex configuration in the system. In Fig. 3(b) we show our MC result for the hysteresis curve for a nanodot of size $r/a=20$, $B/J=3.0$, and $\xi=3$. We can observe two central plateaus with a small inclination. These plateaus in the hysteresis curves can be associated to the presence of vortex configurations. The appearing of the vortex excitation decrease the magnetization of the pure ferromagnetic arrangement. This clearly shows that the vortex excitation appears as the lower energy excitation in the system with no need of introducing it in an artificial way. In Fig. 3(c) MC results are shown for the hysteresis curve for a nanodot of size $r/a=20$, $B/J=3.0$, and $\xi=4$. We observe that the slope of the central plateau is smaller than the former case in a clear indication that the energy of the configuration decreases, thus it becomes more stable. Our MC simulation results is in excellent agreement with the experimental results of Ref. 8.

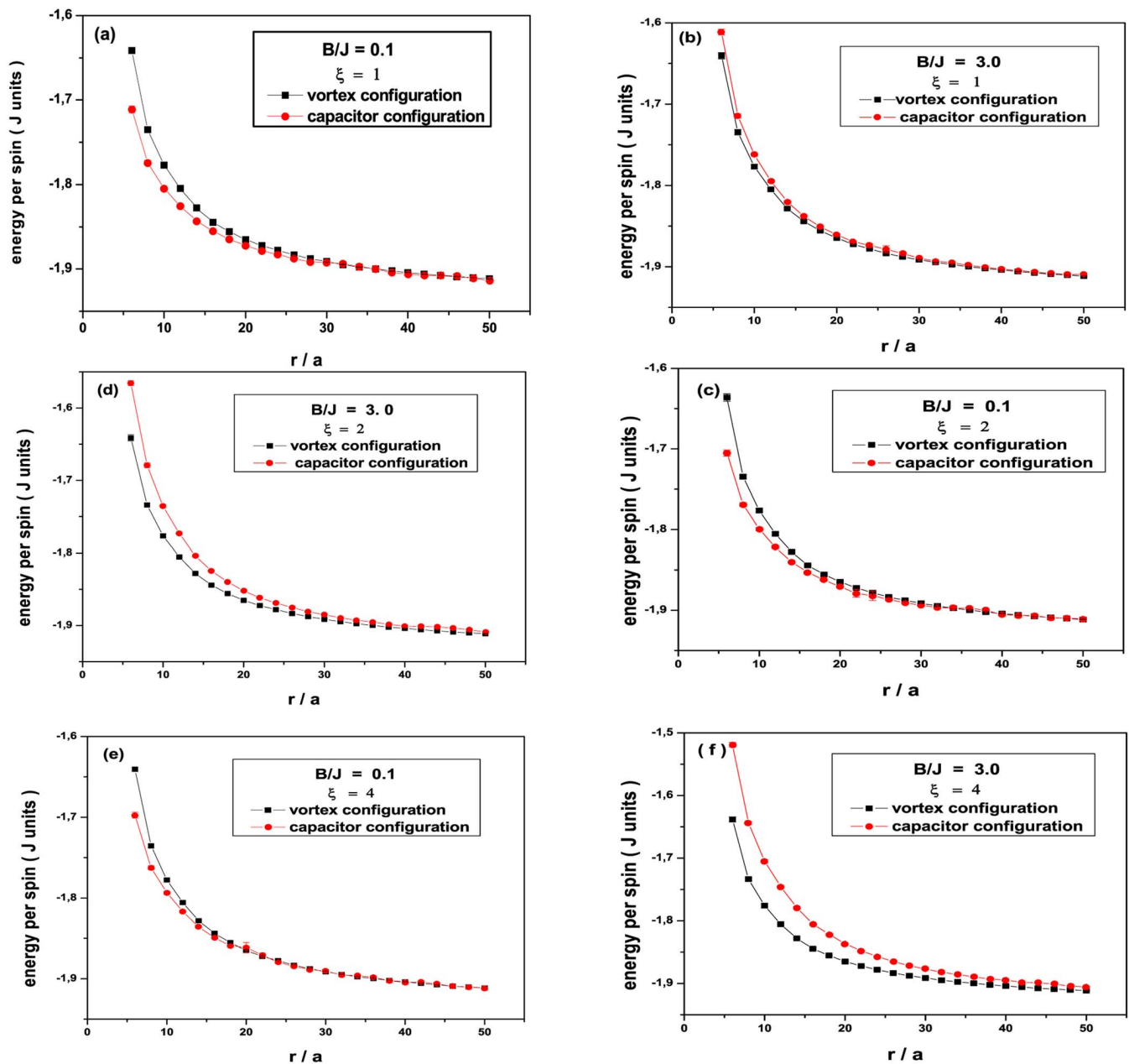

FIG. 2. (Color online) [(a)–(f)] A sequence of plots of the energy of the vortex and capacitor configurations for several values of the surface anisotropy and skin depth ($\xi$), as a function of the nanodot size, at temperature $T/J=0.1$.

In the Figs. 4(a)–4(c) the spin configurations in the nano-dot are shown corresponding to some points of the plateau, indicated with numbers from 1–3, as shown Fig. 3(b).

## IV. CONCLUSIONS

In a recent work Kireev and Ivanov[21] studied the ground and metastable state of a model for magnetic nanodots. They have considered a simple microscopic 2D anisotropic

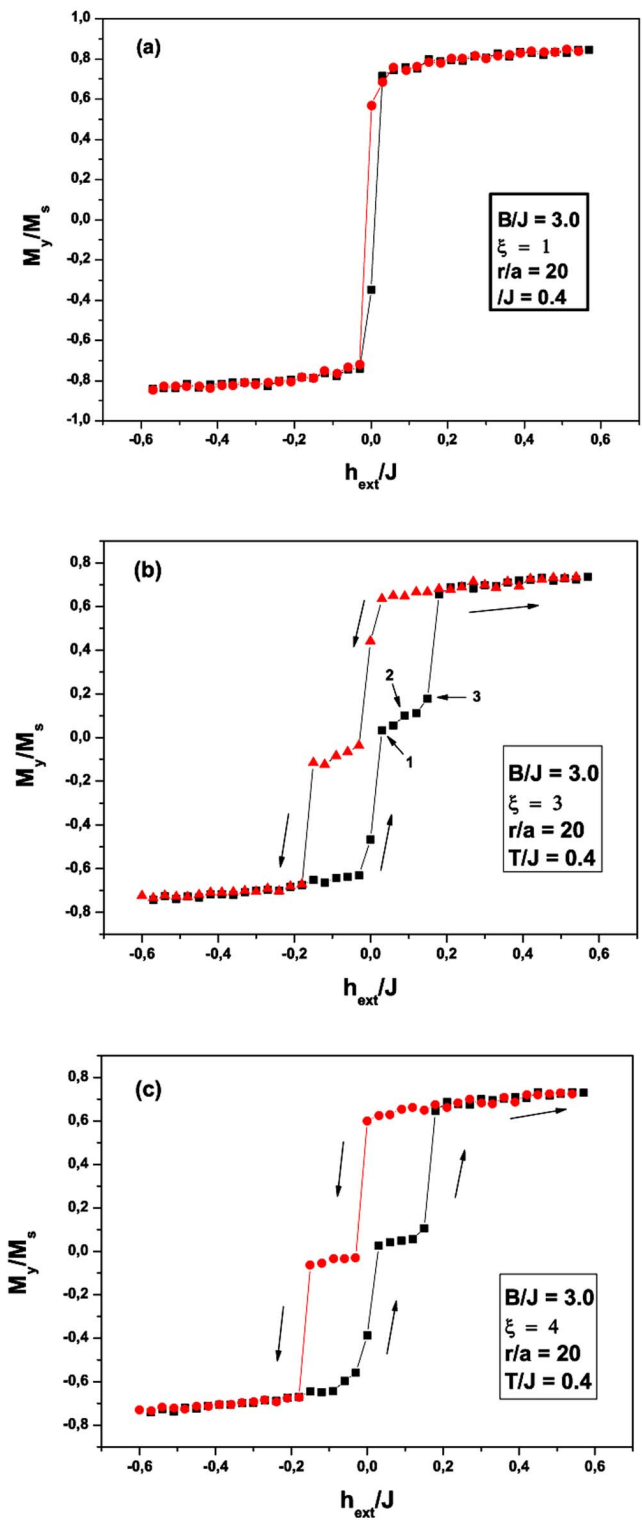

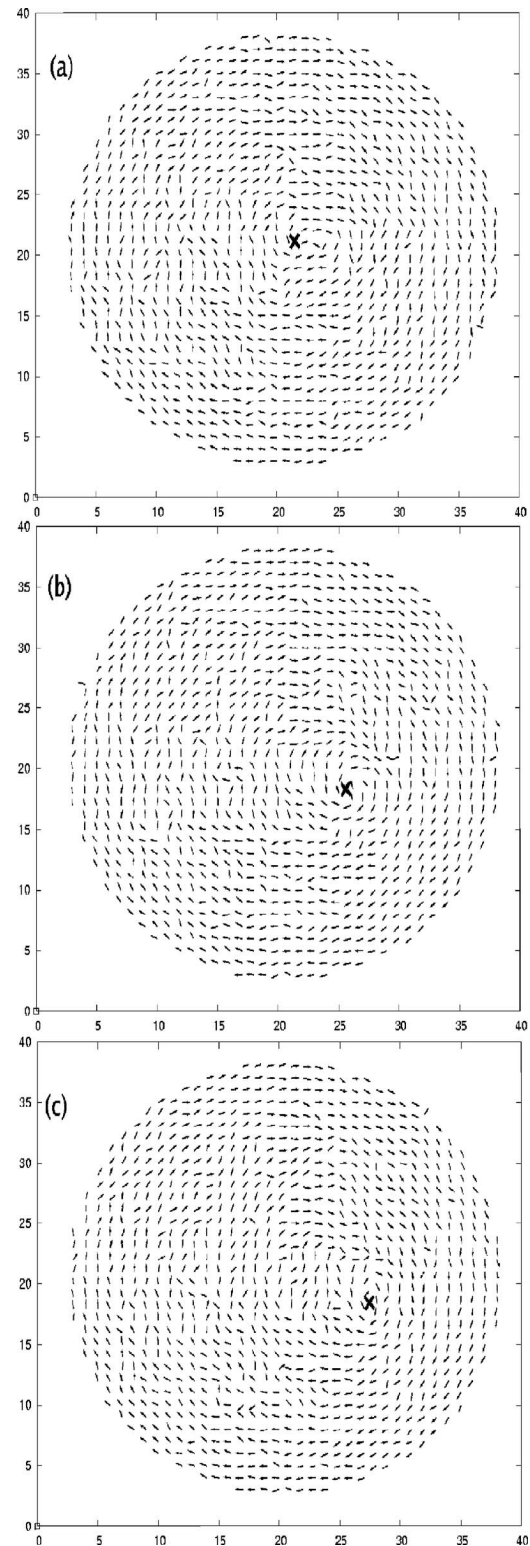

FIG. 3. (Color online) (a) Hysteresis curve for a nanodot of size $r/a=20$, $B/J=3.0$, $\xi=1$, and temperature $T=0.4J$. (b) Hysteresis curve for a nanodot of size $r/a=20$, $B/J=3.0$, $\xi=3$, and temperature $T=0.4J$. (c) Hysteresis curve for a nanodot of size $r/a=20$, $B/J=3.0$, $\xi=4$, and temperature $T=0.4J$. The plateaus can be associated to the presence of vortex configurations.

FIG. 4. [(a)–(c)] A sequence of plots of the spin configurations in the nanodot corresponding to the points 1–3, respectively, of the plateau in the hysteresis curves [Fig. 3(b)]. The values for nanodot size, surface anisotropy, and skin depth are, respectively, $r/a=20$, $B/J=3.0$, and $\xi=3$. The symbol **X** shows the vortex position.

Heisenberg model, with an additional term to describe the anisotropy due to the presence of the surface. They have shown that in the continuous limit the ordered solutions (half vortex) is energetically preferable than the vortex solution. However, experimental results show the existence of stable low energy vortex state at finite temperature in magnetic nanodots. In this work we have considered contribution to the energy of the system due to thickness of the edge which has not been considered in previous works. Introducing a term that controls the energy associated to the edge depth (skin depth), we have performed a careful Monte Carlo simulation of the model. Our results clearly show that the thickness of the edge has a considerable influence in the appearing of a stable vortex state configuration. We have obtained the hysteresis curve for several values of the surface anisotropy (*B*) and skin depth $\xi$. The results we have obtained are in excellent agreement with experimental data.

According to these results, we conclude that the surface depth is an important ingredient to understand the vortex nucleation on magnetic nanodots at low temperature.

## ACKNOWLEDGMENTS

This work was partially supported by CNPq and FAPEMIG (Brazilian agencies). Numerical work was done at the Laboratório de Computação e Simulação do Departamento de Física da UFJF.